\documentclass[twocolumn,showpacs,preprintnumbers,amsmath,amssymb,aps]{revtex4}

\usepackage{amssymb}
\usepackage{amsmath,amssymb,graphicx}
\usepackage{graphicx}
\usepackage{dcolumn}
\usepackage{bm}

\begin{document}


\title{ QFT results for neutrino oscillations and New Physics}%

\author{D. Delepine}
 \email{delepine@fisica.ugto.mx}
\affiliation{Physics department, Division de Ciencias e Ingenieras,  Universidad de Guanajuato, Campus Leon,
 C.P. 37150, Le\'on, Guanajuato, M\'exico.}%
\author{Vannia Gonzalez Macias}
 \email{vanniagm@yahoo.com}
\affiliation{Physics department, Division de Ciencias e Ingenieras,  Universidad de Guanajuato, Campus Leon,
 C.P. 37150, Le\'on, Guanajuato, M\'exico.}%
\author{Shaaban Khalil}
 \email{Shaaban.Khalil@bue.edu.eg}
\affiliation{Centre for Theoretical Physics, The British University in Egypt, El Sherouk City,
Postal No, 11837, P.O. Box 43, Egypt}%
\author{G. Lopez Castro}
 \email{glopez@fis.cinvestav.mx}
\affiliation{Departamento de Fisica, Cinvestav, Apartado Postal 14-740, 07000 Mexico D.F., Mexico}%

\date{\today}

\begin{abstract}
The CP asymmetry in neutrino oscillations, assuming new physics at
production and/or detection processes, is analyzed. We compute this CP
asymmetry using the standard quantum field theory within a general new
physics scenario that may generate new sources of CP and flavor
violation. Well
known results for the CP asymmetry are reproduced in the case of V -A
operators, and additional contributions from new physics operators are
derived. We apply this formalism to SUSY extensions of the Standard
Model where the contributions from new operators could produce a CP
asymmetry observable in the next generation of neutrino experiments.
\end{abstract}

\maketitle

\section{Introduction.}

Since the experimental results implying that the neutrinos are
massive \cite{Fukuda:1998mi}, it is expected to have CP violation
phases in the leptonic sector. However, the situation in the
lepton{\bf ic} sector is very different from the quark sector where the
CP
violation is clearly established in $K$ and $B$ mesons physics.
The only evidence for flavor violation in the leptonic sector
comes from neutrino oscillations and there is, so far, no
confirmation for CP violation in leptonic decays. Hence, measuring
any CP asymmetry in neutrino oscillation will open a new window
to study CP violation and related problems as leptogenesis or
CP violation in $\tau$ decays.  New physics beyond the
Standard Model (SM), like low energy supersymmetry, may be probed
at the LHC and consequently new sources of CP and flavor
violation can be expected. These new sources of CP and lepton
flavor violation, also classified as  non-standard interactions
(NSI),  could give important contributions to the CP
asymmetry in neutrino oscillation (see
ref.\cite{GonzalezGarcia:2001mp,Ota:2001pw,Ota:2005et,Kopp:2007ne}).

In addition, a proper procedure to take into account these non-standard
interactions is important as neutrino experiments are reaching a
high level of accuracy. Even if neutrino masses and lepton mixing matrices
solve the solar and atmospheric neutrinos anomalies, it is clear that
NSI could affect the oscillations parameters as determined by the next
generation of neutrino experiments.  Effects of NSI  have
been widely studied, considering their contributions to
solar and atmospheric neutrino problems, in neutrinos factories,
in conventional and upgraded neutrino beta beams, $e^+e^-$ colliders,
neutrino-electron and neutrino-nucleus scattering and in many other
aspects of neutrino physics
\cite{Kopp:2007ne,Maltoni:2008mu,Kopp:2007mi,Ohlsson:2008gx,Davidson:2003ha,Huber:2002bi,Barranco:2005ps,Barranco:2007tz,Berezhiani:2001rs}.

 The interpretation of a possible CP asymmetry in neutrino
oscillation due to the SM or NSI is still an open question. The
following two
hypothesis are usually considered in the analysis of CP
asymmetry in neutrino oscillations: $(i)$ The probability of a
process associated to neutrino oscillation can be factorized into
three independent parts: the production process, the oscillation
probability and the detection cross section. $(ii)$ The CP
asymmetry in this process is due to the CP violating phase in the
lepton mixing matrix. In a pioneering work, the authors of
Ref. \cite{GonzalezGarcia:2001mp} have studied CP violating effects
due to contributions from new neutrino interactions in the
production and/or detection processes in neutrino oscillation
experiments.  However, only corrections to the V-A SM
charged current interactions were considered
\cite{GonzalezGarcia:2001mp}. In Ref.\cite{Ota:2001pw}, the (V-A)(V-A) and
(V-A)(V+A) operators associated to muon decays, but not to the pion decay, have been considered.
In muon decays, the interference between SM and New Physics of (V-A)(V+A)
operators is suppressed by $m_e/m_{\mu}$ \footnote{ In case of pion decays
the authors of Ref. \cite{Ota:2001pw} have mentionned that it is possible to write the initial
neutrino state as a linear combination of flavour states but no
explicit forms were given for the operators producing these New Flavour
Interactions and how to connect them to  $|\epsilon_{ij}^s>$.}.
In Ref. \cite{Grimus:1996av}, the quantum field theory formalism has
been  used to describe neutrino oscillations but New Flavour
Interactions were not taken into account.

The goal of this paper is to go beyond this approximation and to
propose a generic framework based on quantum field theory to get a
simple expression for the CP asymmetry without imposing any
assumptions on the operators generated by New Physics. We shall show that in such a case,
new contributions to the CP asymmetry appear and it could be
important to take them into account once we want to constraint new
physics using experimental data. We shall illustrate this in the
case of the supersymmetric extension of the Standard Model but it is clear that
our treatment is valid beyond that and it can be applied to NSI
effects in all neutrino experiments.

\section{General Formalism}

Let us
start by giving a sketch of the idea of the present work.
Let us consider a virtual neutrino  that is produced at the space-time
location $(x,t)$, travels to $(x^{'},t^{'})$ and is detected there
because it interacts with a target producing a charged
lepton $l$. For definiteness, we illustrate this process with the
production of the neutrino in $\pi^+$ decay and its later detection via
its weak interaction with a target nucleon $N$ (see Figure 1):
\begin{eqnarray}
\pi (p_1)\rightarrow \mu^{+}(p_2)&+ &\nu_{\mu}^s(p) \nonumber \\
&\hookrightarrow & \nu_{l}^d(p) + N(p_N) \rightarrow N'(p_{N^{'}})+l(p_l)\ . \nonumber
\end{eqnarray}
 We shall call $\nu_{\mu,l}^{s,d}$ state, respectively  the neutrino which is produced at source in conjunction of a $\mu^+$ and the neutrino which is detected through the detection of a charged lepton of flavour $l$. These effective states are not necessary of $\mu$ or $l$ flavour once NSI are introduced.

\begin{figure}
  \includegraphics[width=6cm]{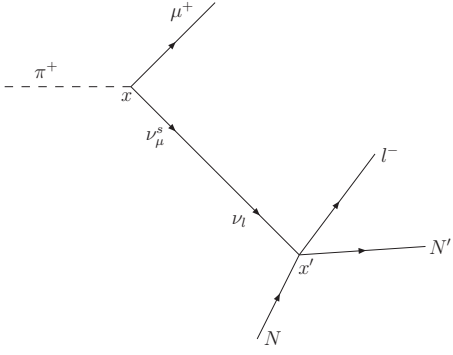}\\
  \caption{Feynman diagram of the process $\pi (p_1)\rightarrow \mu^{+}(p_2)+ \nu_{\mu}^s(p)$ followed by the detection process:$\nu_l(p) + N(p_N) \rightarrow N'(p_{N^{'}})+l(p_l)$. It is important to recall that the flavour of neutrino intermediary state is unobservable.}\label{fig}
\end{figure}

Energy-momentum conservation at the production and detection vertices
requires $p_1=p_2+p$ and $p+p_N=p_{N'}+p_l$. Weak interactions in the SM
acting at the production and detection vertices conserve lepton number
and this process is  interpreted as a  flavor change: a
$\mu$-neutrino is
transformed  into a $l$-neutrino due to oscillation.
However, in the presence of new physics flavor violating weak
interactions can occur, and the flavor identification of the neutrino at
the production and/or decay locations via its associated charged lepton
is not unique anymore.

This process is relevant
for neutrino factories and long-baseline accelerator
experiments such as superbeams. Pions are produced at the source
when a high-energy proton beam ($\sim 10^{2} \,\mathrm{GeV}$)
hits a target (made of solid dense material e.g. Be, Al, Ca,
etc...).
Superbeam neutrino sources are mainly from pion decays while neutrino
factories sources are mainly from muons.


 If we assume that light neutrinos are left-handed, Lepton Flavor (LF)-violating
semi-leptonic interactions can be described by the following effective
Hamiltonian:

\begin{eqnarray}
{\cal{H}}&=&2\sqrt{2}G_FV_{ud}\left\{  C_1^{k}
\left(\overline{\mu}\gamma^{\alpha}P_L\nu_{k}\right)\left(\overline{u}\gamma_{\alpha}P_Ld
\right) \right. \nonumber \\
& & + C_2^{k} \left(\overline{\mu}\gamma^{\alpha}P_L\nu_{k}\right) \left(\overline{u}\gamma_{\alpha}P_Rd \right)\nonumber \\
& & + C_3^{k} \left(\overline{\mu}P_L\nu_{k}\right) \left(\overline{u}P_Ld\right) \nonumber \\
& & + C_4^{k} \left(\overline{\mu}P_L\nu_{k}\right) \left(\overline{u}P_Rd \right) \nonumber \\
& & \left.  + C_{5R}^{k} \left( \overline{\mu}\sigma_{\alpha \beta
}P_L\nu_{k}\right)  \left(\overline{u}\sigma^{\alpha \beta}P_Rd\right)
\right. \nonumber\\
& & \left. + C_{5L}^{k} \left( \overline{\mu}\sigma_{\alpha \beta
}P_L\nu_{k}\right) \left(\overline{u}\sigma^{\alpha \beta}P_Ld\right)
\right\}\ ,
\end{eqnarray}
where $k$ runs over the three leptonic flavors and $P_{R,L}=(1\pm
\gamma_5)/2$.

In the following and for simplicity, we  consider the case where LF violation occurs only
at the  $\pi^+$ decay  vertex. It is straighforward to include the
effects of such New Physics  at the detection vertex using this
formalism. Note that the tensor currents proportional to
the $C_{5L(R)}^{k}$ Wilson coefficients will not contribute to $\pi^+$
decay because it is not possible to generate an antisymmetric
tensor from the pion momentum alone. Thus, the only non-vanishing
hadronic matrix elements at the production vertex are:
\begin{eqnarray}
\langle 0|\overline{d}\gamma^{\mu}\gamma_{5}u|\pi^{+}\rangle&=&
if_{\pi}p_{\pi}^{\mu} \\
 \langle 0|\overline{d}\gamma_5u|\pi^{+}\rangle&=&\frac{
-if_{\pi}m_{\pi}^2}{m_u+m_d} \ ,
\end{eqnarray}
where $f_{\pi}=130$ MeV is the pion decay constant and $m_{u,d}$
denote the light quark masses.

Using the relation
$\nu_k=\sum U_{k\alpha} \nu_{\alpha}$ between flavor $k$ and mass
$\alpha$ neutrino eigenstates, we get the following amplitude for
$\pi^+$ decay:
\begin{equation}
\langle \mu^{+} \nu_{k}|{\cal{H}}|\pi^{+} \rangle =
\bar{u}(p) O^{k} v(p_2)\ ,
\end{equation}
 where:
\begin{eqnarray}
{\cal{O}}^{k}& \equiv & \frac{G_F}{\sqrt{2}}V_{ud} (1+\gamma_5)f_{\pi}
\left(
\frac{-im_{\pi}^2}{m_u+m_d}(C_{3}^k-C_{4}^{k}) \right. \nonumber \\
& & \left.-i(C_1^k-C_2^k){\not p}_{\pi} \right)
\end{eqnarray}
with ${\not p}_{\pi}\equiv \gamma_{\mu}p_{\pi}^{\mu}$.

Now consider a neutrino of flavor $k$ that is produced in $\pi^+$ decay
via LF-violating interactions and is detected with flavor $l$ at a later
time via (LF-conserving) charged current scattering off the nucleon $N$
(Fig. 1). In  Quantum Field Formalism, the S-matrix amplitude for the evolution for the system from initial state is given by:
 \begin{eqnarray}
T_{\nu_{\mu}^s-\nu_{l}}&=& \int d^4x d^4x^{'}\sum_k
e^{i(p_l-p_N+p_{N^{'}})\cdot x^{'}}\frac{G_F V_{ud}}{\sqrt{2}}
(J_{NN^{'}})_{\mu} \nonumber \\
& & \overline{u}_l(p_l)\gamma^{\mu}(1-\gamma_5)
\Delta_{\nu}^{lk}(x^{'}-x){\cal{O}}^kv(p_2) e^{i(p_2-p_1)\cdot x} \label{1}
\nonumber
\end{eqnarray}
where $\Delta_{\nu}^{lk}(x^{'}-x)$ basically describes oscillation of
neutrinos during its propagation. It is important to stress that we never introduce Energy-momentum eigenstates to describe the neutrinos as in our formalism the neutrinos appear as virtual particles.
In terms of massive neutrino propagators we can write
\begin{eqnarray}
\Delta_{\nu}^{lk}(x^{'}-x)&=&\sum_i U_{li}U_{ki}^{*} \int
\frac{d^4p}{(2\pi)^4}e^{-ip(x^{'}-x)}\frac{i}{{\not p}-m_{\nu_{i}}+
i\epsilon}\nonumber
\end{eqnarray}
 where $U_{li}$ are the usual elements of the $U_{MNS}$ mixing
matrix. We write this propagator in a more convenient form by
integrating upon the time component of the four-momentum
\cite{Beuthe:1997fu}:
\begin{eqnarray}
&&\Delta_{\nu}^{lk}(x^{'}-x)= \sum_i U_{li}U_{ki}^{*} \int\frac{d^3p}{(2\pi)^3}  \\
& & \times \left( \frac{e^{-iE(t^{'}-t)}e^{i\overrightarrow{p}(\overrightarrow{x^{'}}-
\overrightarrow{x})}}{2E_{\nu_i}} (E_{\nu_i}.\gamma^0-\overrightarrow{p}.\overrightarrow{\gamma}+m_{\nu_{i}}) \theta(t^{'}-t) \right. \nonumber \\
& & \left.+\frac{e^{iE(t^{'}-t)}e^{i\overrightarrow{p}(\overrightarrow{x}'
-\overrightarrow{x})}}{2E_{\nu_i}} (-E_{\nu_i}.\gamma^0-\overrightarrow{p}.\overrightarrow{\gamma}+m_{\nu_{i}})\theta(t-t^{'})\right) \nonumber
\end{eqnarray}
 where $E_{\nu_i}=\sqrt{\vec{p}^2+m_{\nu_i}^2}$. As usual, we interpret the first term as neutrinos  propagating forwards in
time and the second as anti-neutrinos propagating backwards in time.

Thus, by keeping only the first term of the propagator in Eq. (6)
we get the amplitude
\begin{eqnarray}
T_{\nu_{\mu}^s-\nu_{l}}&=& i\int
\frac{d\tau}{2E_{\nu_i}} e^{iP_{F}^0\tau} (2 \pi)^4
\delta^4(P_{F}+p_2-p_1) \nonumber \\
&&\  \times \frac{G_F V_{ud}}{\sqrt{2}}
(J_{NN^{'}})_{\mu} \sum_{i,k} U_{li}U_{ki}^{*} \overline{u}_l(p_l)
\gamma^{\mu}(1-\gamma_5) \nonumber \\
&&\ \times e^{-i\tau(E_{\nu_i})}(\not P_{F}+
m_{\nu_i}){\cal{O}}^kv(p_2)
\end{eqnarray}
with $\tau =(t^{'}-t)>0$  is the time elapsed from the production to the detection
space-time locations of neutrinos and $P_{F}\equiv p_l-p_N+p_{N^{'}}$.
Equivalently, the time-dependent  amplitude from initial to final states is the integrand of eq.(7):
\begin{eqnarray}
T_{\nu_{\mu}^s-\nu_{l}}(\tau)&=&  (2 \pi)^4 \delta^4(P_{F}+p_2-p_1) (G_F V_{ud})^2 (J_{NN^{'}})_{\mu}\nonumber \\
&&\times f_{\pi}\sum_k\overline{u}_l(p_l) \gamma^{\mu}(1-\gamma_5)
(\not p_l-\not p_N+\not p_{N^{'}}) \nonumber \\
&& \times  \left(m_{\pi} A_k^{*}+B_k^{*} {\not p_{\pi}} \right)v(p_2) e^{iP_{F}^{0}\tau} \nonumber \\
&& \times \sum_i  U_{li}U_{ki}^{*}\frac{e^{-i\tau(E_{\nu_i})}}{2E_{\nu_i}}
\ ,
\end{eqnarray}
where
\begin{eqnarray}
B_k&\equiv &  (C_1^{k*}-C_2^{k*}] \ , \\
A_k&\equiv & \frac{ m_{\pi}}{m_u+m_d}(C_3^{k*}-C_4^{k*})\ ,
\end{eqnarray}

The time evolution amplitude for the corresponding CP-conjugate process which correspond to the observation at source of a $\mu^-$ and detection of a $l^+$
is given by:
\begin{eqnarray}
T_{\overline{\nu}^s_{\mu}-\overline{\nu}_{l}}(\tau)&=&  (2 \pi)^4 \delta^4(P_{F}+p_2-p_1) (G_F V_{ud})^2 (J_{NN^{'}})_{\mu}\nonumber \\
&&\times f_{\pi} \sum_k\overline{v}_l(p_l) \gamma^{\mu}(1-\gamma_5)
(\not p_l-\not p_N+\not p_{N^{'}}) \nonumber \\
&& \times  \left(m_{\pi} A_k+B_k {\not p_{\pi}}\right)u(p_2) e^{iP_{F}^{0}\tau}\nonumber \\
&& \times \sum_i  U_{li}^{*}U_{ki} \frac{e^{-i\tau(E_{\nu_i})}}{2E_{\nu_i}}
\ .
\end{eqnarray}

In order to get the CP asymmetry in the general case, let us choose the
following form of the nucleon weak vertex:
\begin{equation}
(J_{NN^{'}})_{\mu}=\overline{u}_{N^{'}}(p_{N^{'}})\gamma_{\mu}(g_V+g_A \gamma_5) u_{N}(p_{N})
\end{equation}
with $g_V=g_V(q^2=0)=1$ and $g_A= g_A(q^2=0)\approx -1.27$
\cite{Yao:2006px}.
Under these approximations, one has:
\begin{eqnarray}
|T_{\nu_{\mu}^s-\nu_{l}}(t)|^2& \equiv &\sum_{q,k}(B_q^* m_{\mu}- m_{\pi}A_q^*)(B_k m_{\mu}- m_{\pi}A_k)  \nonumber \\
&&U_{li}^* U_{ki} U_{lj}
U_{qj}^* e^{i\tau (E_{\nu_i}-E_{\nu_j})} F(P,M) \ ,
\end{eqnarray}
where  $F(P,M)$ is a kinematical function that depends on masses and momenta
of external particles  but not on neutrino flavour and will drop in the $a_{CP}(\tau)$ asymmetry defined as:
\begin{eqnarray}
a_{CP}(\tau)&=&\frac{|T_{\nu_{\mu}^s-\nu_{l}}(\tau)|^2-|T_{\overline{\nu}_{\mu}^s
-\overline{\nu}_{l}}(\tau)|^2}{|T_{\nu_{\mu}^s-\nu_{l}}(\tau)|^2+
|T_{\overline{\nu}_{\mu}^s-\overline{\nu}_{l}}(\tau)|^2} \\
&\equiv & \frac{N(\tau)}{D(\tau)}
\end{eqnarray}

\section{$CP$ asymmetry from $C_1$ Wilson coefficient ((V-A)(V-A) operator)}.

In the usual formalism developed by refs.
\cite{GonzalezGarcia:2001mp}\cite{Ota:2001pw}
one considers only New Physics corrections due to $
C_1^k= C_{SM}(\delta_{k\mu}+\epsilon_{\mu k})$.
In this approximation, one gets for the amplitude
\begin{eqnarray}
&&T_{{\nu}_{\mu}^s-{\nu}_{l}}(\tau)=  (2 \pi)^4  e^{iP_{F}^{0}\tau}\delta^4(P_{F}+p_2-p_1) (G_F V_{ud})^2 (J_{NN^{'}})_{\mu}\nonumber \\
&&\times  f_{\pi} C_{SM} \overline{u}_l(p_l) \gamma^{\mu}(1-\gamma_5)
(\not p_l-\not p_N+\not p_{N^{'}}){\not p_{\pi}}v(p_2) \nonumber \\
&& \times \sum_{i,k}  U_{li}U_{ki}^{*} \frac{e^{-i\tau(E_{\nu_i})}}{2E_{\nu_i}} (\delta_{k\mu}+\epsilon_{\mu k})
\end{eqnarray}
Once taking the $ |T_{\nu_{\mu}^s-\nu_{l}}(\tau)|^2 $, the first two lines of the previous equation will give us the kinematical part of the process which is common to all neutrino flavor and the last line will give, at first order in neutrino masses, the usual neutrino oscillation results including Non Standard Interactions. It is interesting to note the equivalence of our expression with the approach done in ref.\cite{GonzalezGarcia:2001mp} where they defined:
\begin{equation}
|\nu_{\mu}^{s}> = \sum_{i} \left( U_{ei}^{*} \epsilon_{\mu e} + U_{\mu i}^{*} (1+\epsilon_{\mu \mu})+U_{\tau i}^{*} \epsilon_{\mu \tau}\right) |\nu_{i}>
\end{equation}
where $|\nu_{i}>$ are the neutrino mass eigenstates and $|\nu_{\mu}^{s}>$ is the initial flavour state produced at neutrino source. Computing $|\langle \nu_l|\nu_{\mu }^s(\tau)\rangle|^2$, one obtains the neutrino oscillation part of  $ |T_{\nu_{\mu}^s-\nu_{l}}(t)|^2 $ once assuming that $1/2E_{\nu_i} = (1/2E_{\nu}) (1+ {\cal{O}}(m_{\nu_i}^2))$ and keeping the first term of the expansion in $m_{\nu_i}^2$.

\section{$CP$ asymmetry from $C_{3,4}$ Wilson coefficients}
The interesting results of this formalism is that naturally all NSI can be taken into account without making any a priori assumptions. In particular, one can easily include the effects of new CP violating phases and flavour-violating interactions from $C_{3,4}$ Wilson coefficients which involve scalar and pseudoscalar density operators.
 As an example, let us assume  that the only sources of CP
violating phases are coming from the scalar operators $C_{3,4}^k$
(equivalently a
relative weak phase between $A_k$ and $B_k$, different from $n\pi$), so
the numerator of the  CP asymmetry will be proportional to
\begin{eqnarray}
N(\tau) &\propto & Re\left(\sum_{i,j,k,q} A_k B_q^* U_{li}^* U_{ki}
U_{lj} U_{qj}^* e^{-i\tau (E_{\nu_i}-E_{\nu_j})} \right) \ \nonumber\\
&& - Re\left(\sum_{i,j,k,q} A_k^* B_q U_{li} U_{ki}^*
U_{lj}^* U_{qj} e^{-i\tau(E_{\nu_i}-E_{\nu_j})} \right) \ .\nonumber
\end{eqnarray}
The Standard model contributes only to $C_1^k =C_{SM}\delta_{k \mu}+$ corrections from New Physics
(in our convention, $C_{SM}=1$) and $C_{3,4}$ are produced by New Physics.  At first order in new physics, one has to replace $B_q$ by $\delta_{q \mu}$ in the CP asymmetry as previously defined.

It is interesting to note that we could get this result by assuming from
the beginning that
\begin{eqnarray}
a_{CP}(\tau)&=&\frac{|\langle \nu_l | \nu_{\mu}^s(\tau) \rangle|^2-|\langle \overline{\nu_l} |\overline{ \nu_{\mu}^s}(\tau) \rangle|^2}{|\langle \nu_l | \nu_{\mu}^s(\tau) \rangle|^2+|\langle \overline{\nu_l} |\overline{ \nu_{\mu}^s}(\tau) \rangle|^2}
\end{eqnarray}
where one defines the
\begin{equation}
|\nu_{\mu}^{s}> = \sum_{i} \left( U_{ei}^{*} \epsilon_{\mu e} + U_{\mu i}^{*} (1+\epsilon_{\mu \mu})+U_{\tau i}^{*} \epsilon_{\mu \tau}\right) |\nu_{i}>
\end{equation}
with
$\epsilon_{k\mu}\equiv \frac{A_k}{C_{SM}}$.  It is worth mentioning that using the QFT formalism, all $\epsilon_{ij}^{s,d}$ are expressed in terms of the Wilson coefficients and once the $\epsilon_{ij}^{s,d}$ are defined, the analysis considered in Ref.\cite{GonzalezGarcia:2001mp,Ota:2001pw} can be easily implemented. As one can see from eq.(11), the New Physics is enhanced by a factor {\bf $m_{\pi}/(m_u+m_d) \approx 15$}  which multiplies the Wilson coefficients $C_{3,4}^{k}$ as mentionned previously
in refs.\cite{Herczeg:1995kd}\cite{Vainshtein:1975sv}.
It is clear that the limit on helicity of the muon in pion decays
\cite{Fetscher:1984da}\cite{Yao:2006px} will constraint the
contribution of $C_{3,4}$ not to be bigger than a few percent.
It is important to note that most of the constraints on $\epsilon_{ij}$ come
 from four lepton Fermi Operators (see ref.\cite{Davidson:2003ha}) as in
$ \mu \rightarrow eee$ or rare muon decays as $\mu \rightarrow e
\gamma$; also some lepton flavour violating tau decays impose very
strong constraints on $\epsilon_{\mu i}$ with $i=e $ or $\tau$.  In
the case of operators which contribute  to  pion decays, one should
note that
these operators can be written as the product of a leptonic current
and a hadronic current. Thus, the Wilson coefficents in both cases
can not be related to each other unless they can be factorized into the
product of a leptonic and a hadronic part and if CP and flavor
violations occur only in the leptonic vertices. This procedure is model-dependent
and should be verified in each specific model of New Flavour Interactions.

\section{ Application to SUSY models}

Let us apply this formalism to the usual Minimal Supersymmetric
extension of the SM.
Hence, as an application, we calculate the supersymmetric (SUSY) contributions to the dominant pion decay mode
$\pi^{-}\rightarrow \mu^{-}\bar{\nu_{\mu}}$ (or its charge conjugate) \cite{fayet,fayet2,ffayet} through the Wilson coefficients and focus our attention on the SUSY constributions to scalars and pseudoscalars  Wilson coefficients ($C_3$ and $C_4$).

\begin{figure}
  \includegraphics[width=8cm]{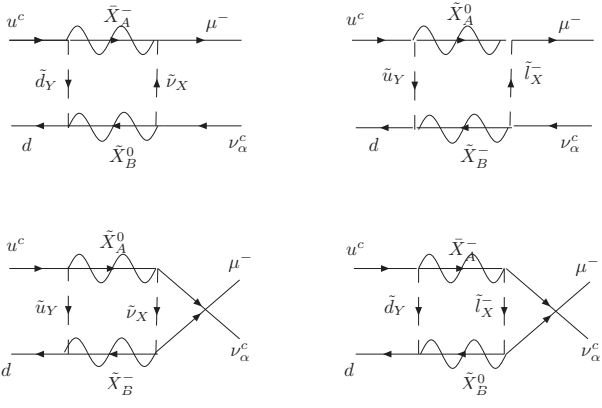}\\
  \caption{Feynman diagrams of dominant SUSY contribution to $\pi^- \to \mu^- + \nu^c_\alpha$.}\label{fig2}
\end{figure}

The effective lagrangian contributing to the $\nu_{\mu} \rightarrow
\nu_{k}$ flavor change at the neutrino source, is given by
\begin{eqnarray}
{\cal H}_{eff}=2 \sqrt{2} G_{F} V_{ud} \sum_{j}C_{j}^{k}
{\cal{O}}_{j}^{k} \ ,\nonumber
\end{eqnarray}
where $C_{j}^{k}$ are the dimensionless Wilson coefficients and
${\cal{O}}_{j}^{k}$ are the relevant local operators at low energy
scale as defined in Eq.(2).
The leading order contributions to $C_3$ and $C_4$ under the experimental constraints for
the Wilson coefficients induced by SUSY are explicitly shown in the
expressions below.
These describe box type diagrams of chargino-neutralino exchanges.
Other SUSY contributions (vertex corrections) are suppressed by the
Yukawa couplings of light leptons.
 In fig ({\ref{fig2}}), the Feynman diagrams of dominant SUSY contribution to the pion decay, $\pi^- \to \mu^- + \nu^c_\alpha$, are represented.

To simplify the expression, we shall
assume that the squark and slepton masses are degenerated ($\tilde{m}_f
\simeq m_{\tilde{u},\tilde{d}} \simeq m_{\tilde{l},\tilde{\nu}}$)\cite{Barbieri:1995tw}:
\begin{eqnarray}
C_{3}^{k}&=& -\sum_{X,Y,A,B}
\frac{g^{4}}{2\sqrt{2}G_{F}V_{ud}}(\frac{4}{3}\tan\theta_{W}N^{*}_{B1}) (\sqrt{2}\tan \theta_{W} N^{*}_{B1})\nonumber\\
&& \times  \left( -V^{*}_{A1}U^{CKM}_{Y,1}+\frac{m^{u}_{Y}}{\sqrt{2}M_{W}\sin\beta}V^{*}_{A2}U^{CKM}_{Y,1} \right)\nonumber\\
&& \times \left(-U_{A1}U^{MNS}_{X,k}  +\frac{m^{l}_{X}}{\sqrt{2}M_{W}\cos \beta}U_{A2}U^{MNS}_{X,k}\right) \nonumber \\
&&  \times M_{\tilde{X}^{0}_{B}}M_{\tilde{X}^{-}_{A}}  J(x_a,x_b )
\end{eqnarray}
\begin{eqnarray}
C_{4}^{k}&=&  \sum_{X,Y,A,B} \frac{g^{4}}{2\sqrt{2}V_{ud} G_{F}}(\sqrt{2}\tan \theta_{W}N^{*}_{B1} )(\frac{2}{3}\tan\theta_{W}N_{B1}) \nonumber \\
&&
 \times  \left(-U_{A1}U^{CKM}_{1Y}+\frac{m^{d}_{Y}}{\sqrt{2}M_{W}\cos\beta}U_{A2}U^{CKM}_{1Y}\right) \nonumber \\
 && \times \left(-U_{A1}U^{MNS}_{X,k} +\frac{m^{l}_{X}}{\sqrt{2}M_{W}\cos\beta}U_{A2}U^{MNS}_{X,k}\right) \nonumber\\
 && \times  I(x_a,x_b) \qquad
\end{eqnarray}
where the $N$ and $U,V$ matrices are the matrices which respectively
diagonalize the neutralinos and chargino mass matrices, $g$ is the gauge
weak coupling and $\theta_W $ is the Weinberg angle. The sum on
$X,Y,A,B$ is respectively on the squarks, sleptons, charginos and
neutralinos.
The functions $I(x_{a},x_{b}),J(x_{a},x_{b})$ are given by
\begin{eqnarray}
I(x_{a},x_{b})&=& \frac{1}{16\pi^{2}\tilde{m_{f}}^{2}} \left(\frac{1}{x_a-x_b}\right) \nonumber \\
&& \times \left(\frac{-x_a+x_a^2-x_a^2 \ln x_a}{(1-x_a)^2}- (x_a \rightarrow x_b)\right) \nonumber \\
J(x_{a},x_{b})&=& \frac{1}{16\pi^{2}\tilde{m_{f}}^{4}} \left(\frac{1}{x_a-x_b}\right) \nonumber \\
&& \times \left( \frac{-1+x_a-x_a \ln x_a}{(1-x_a)^2}- (x_a \rightarrow x_b) \right) \nonumber
\end{eqnarray}
and $x_{a,b}$ are defined as
$x_{a}=\frac{\tilde{m}^{2}_{X^{\pm}}}{\tilde{m}^{2}_{f}},\,x_{b}=\frac{\tilde{m}^{2}_{X^{0}}}{\tilde{m}^{2}_{f}}$ .
In order to have non-vanishing asymmetry we should have a
non-vanishing imaginary part of $C_3$ or $C_4$. The complex phase of $C_{3,4}$ can be
due to the matrices that diagonalize the squark mass matrix or the chargino
mass matrix. If we assume that the squark matrices are diagonal, i.e the
diagonalizing matrices are identity, we do not have any source of CP
violation from squark matrices. Then the only remaining source is the phase
of $\mu$ which induces a complex phase in chargino mass matrix and hence $U$ and
$V$ unitary matrices. This phase is strongly
constrained by the neutron electric dipole moment (EDM)\cite{Baker:2006ts,Arnowitt:1990je,Barr:1999hf,Ellis:2008zy,Romalis:2000mg,Kizukuri:1992nj} but it is possible to avoid this constraint if the first generation of squarks are heavy enough and decouple from the two other squark generations.
 Another way to avoid the neutron EDM constraint is to assume
flavor violating structure in the squarks and sleptons mass matrices.
This allows to have new sources of CP violation which could contribute
to any CP violating observables and/or could relax the constraint on
the $\mu$ term coming from neutron EDM limit\cite{Ellis:2008zy}.

\begin{figure}
  \includegraphics[width=8cm]{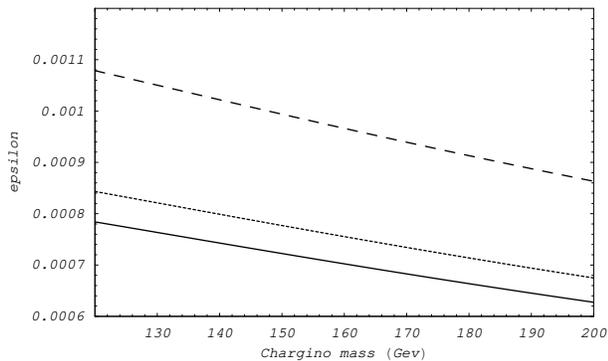}\\
  \caption{Absolute value of $\epsilon_{\mu e}$ (solid line) and $\epsilon_{\mu \tau}$ (dashed and dotted lines) for different choice of $\tan\beta$, respectively $\tan \beta=50$ and $10$. The other parameters are given by $\widetilde{m}_f=100$ GeV and $\widetilde{m}_{\chi^0}=110$ GeV.}\label{fig3}
\end{figure}

Assuming that the CP violating sources comes from the $\mu$ term in the
chargino sector, $\epsilon_{\mu e}$ is given by $C_{3,4}$ times the
enhancement factor $m_{\pi}/(m_u+m_d)$. For typical chargino
and  neutralino masses of order 150 GeV and sfermion masses around
100  GeV, it is possible to obtain $\epsilon_{\mu e}$ as large as
$10^{-3}$.  In order to maximize the absolute values of  $\epsilon$'s, one assumes that charginos and neutralinos have quasi-degenerated masses.  In figure (\ref{fig3}), we present our numerical results for $|\epsilon_{\mu e}|$ (solid line) and $|\epsilon_{\mu \tau}|$ (dashed and dotted lines) as function of the chargino mass, for $\tan \beta=50$ and $\tan \beta=10$. As expected from  the expressions of $C_{3,4}$, $\epsilon_{\mu e}$ is not sensitive to $\tan\beta$ if $\theta_{13}^{MNS} \simeq0$. One should emphasize that the SUSY model presented in this section is the simplest one and we could expect enhancement in flavour violation effects once a non-universal structure is assumed in the soft-SUSY breaking terms.  In this case, the values of $\epsilon_{e\mu}$ or/and $\epsilon_{\mu \tau}$ may be significantly enhanced. A detailed analysis for the CP asymmetry in neutrino oscillation in SUSY model with non-minimual flavor will be considered elsewhere.

 As one can see from Fig. (\ref{fig3}), $\epsilon_{e\mu}$ is typically of order $10^{-3}$. These values lead to CP asymmetry in neutrino oscillation of order $10^{-1}\sim 10^{-2}$, depending of the parameters of the neutrino beam and the size of the experiment baseline \cite{GonzalezGarcia:2001mp}.  Such asymmetries are reachable at next generation of reactor and beam neutrino oscillation experiments \cite{Kopp:2007ne,Maltoni:2008mu,Kopp:2007mi,Ohlsson:2008gx,Davidson:2003ha}.

\section{Conclusions}
To conclude, we proposed a formalism based on Quantum Field Theory where it is possible to include all sources of Non Standard Interactions including new CP- and flavour-violating interactions. We show that using this method, it is straightforward to include the effects of the scalar and pseudoscalar operators densities which appears in any New Physics Models. In the limit where we consider  only (V-A)(V-A) operator, we reproduce the usual results reported in Ref.\cite{GonzalezGarcia:2001mp}. It is important to emphasize that most of the studies on NSI in neutrino physics have been done assuming that New Physics contributions to neutrino interactions is mainly due to corrections to (V-A)(V-A) Wilson coefficients. This approach was well justified when neutrino experiments were not so accurate and other sources of New Physics could be easily neglected. But now neutrino experiments enter in the field of high precision experiments where the effects of NSI coming from (S-P)(S+P) and (S-P)(S-P)  operators could have an impact on the determination of lepton mixing parameters including the CP violating phases.
To illustrate our results, we applied it to SUSY models where we assume that the only CP-violating phases appear in $C_{3,4}$ Wilson coefficients which correspond respectively to (S-P)(S-P) and (S-P)(S+P) operators.  In such a case, we show that with reasonable values for the SUSY parameters, it is possible to generate a CP asymmetry as large as $10^{-1}\sim 10^{-2}$.



\acknowledgments{This work was also partially supported by Conacyt (Mexico) and by PROMEP project. The work of S.K. is supported in part by ICTP project 30 and the Egyptian Academy of Scientific Research and Technology.}

\end{document}